\documentstyle[prl,multicol,epsf,epsfig,aps]{revtex}
\newcommand{\BEQ}{\begin{equation}}
\newcommand{\EEQ}{\end{equation}}
\newcommand{\BEA}{\begin{eqnarray}}
\newcommand{\EEA}{\end{eqnarray}}
\newcommand{\nn}{\nonumber \\}
\renewcommand{\d}{{\rm d}}

\newcommand{\omt}{{\omega_0t\,}}
\newcommand{\omtau}{{\omega_0\tau\,}}

\renewcommand{\nn}{\nonumber}
\newcommand{\eps}{\varepsilon}
\newcommand{\si}{\hat{\sigma}}
\newcommand{\sip}{\hat{\sigma}_{+}}

\newcommand{\om}{\omega}

\newcommand{\e}{\hat{\eta }}
\newcommand{\tr}{{\rm tr}}
\newcommand{\ha}{\hat{a}}
\newcommand{\taut}{\delta_1}

\newcommand{\so}{{\,\sin\theta\,}}

\newcommand{\co}{{\,\cos\theta\,}}

\newcommand{\T}{{\cal T}}
\renewcommand{\H}{{\cal H}}

\newcommand{\half}{\frac{1}{2}}

\newcommand{\Q}{{\cal{Q}}}

\newcommand{\piplus}{\hat{\Pi}_{+}}

\def\dbarrm {{\mathchar'26\mkern-11mu{\rm d}}}                       %
\begin{document}
\draft
\title{
Bath generated work extraction and inversion-free gain in two-level systems
}
\author{A.E. Allahverdyan$^{1,2)}$ and Th.M. Nieuwenhuizen$^{3)}$}
\address{
$^{1)}$ S.Ph.T., CEA Saclay, 91191 Gif-sur-Yvette cedex, France\\
$^{2)}$Yerevan Physics Institute,
Alikhanian Brothers St. 2, Yerevan 375036, Armenia\\
$^{3)}$ Institute for Theoretical Physics, University of Amsterdam\\
Valckenierstraat 65, 1018 XE Amsterdam, The Netherlands}
\maketitle
\begin{abstract}
The spin-boson model, often used in NMR and ESR physics, quantum optics and
spintronics, is considered in a solvable limit to model
a spin one-half particle interacting with a bosonic thermal bath.
By applying external pulses to a non-equilibrium
initial state of the spin, work can be extracted
from the thermalized bath. It occurs on the timescale $\T_2$ inherent
to transversal (`quantum') fluctuations.
The work (partly) arises from heat given off by the
surrounding bath,
while the spin entropy remains constant during a pulse.
This presents a violation of the Clausius inequality
and the Thomson formulation of the second law
(cycles cost work) for the two-level system.

Starting from a fully disordered state, coherence can be induced by employing
the bath. Due to this, a gain from a positive-temperature (inversion-free)
two-level system is shown to be possible.

\end{abstract}
\begin{multicols}{2}
After E.L. Hahn discovered the spin-echo in
NMR physics~\cite{Hahn}, it was soon suspected to
violate the second law \cite{Hahn,Waugh}; for a recent discussion
discussion see e.g. ~\cite{Waugh,balian}. As a precursor to this
question, we shall investigate in this Letter whether single
or double pulses on single spins coupled to a bath
already mark a violation of this law.

Recently we analyzed  the thermodynamics of the
Caldeira-Leggett model for a quantum harmonic oscillator coupled to a
quantum harmonic bath~\cite{ANprl,NAlinw}. At low
temperatures various formulations of the second law are
violated: the Clausius inequality $\dbarrm\Q\le T\d S$ is broken, the rates
of energy dispersion and entropy production can be negative, and certain
cycles
are possible where heat extracted from the bath is fully converted into
work (``perpetuum mobile''). These findings are nevertheless in agreement
with the Thomson formulation of the second law (cycles cost work)
applied to an equilibrium initial distribution, for which an
exact proof exists~\cite{ANthomson}.

The cause of the breakdown the universal thermodynamic picture
is the occurrence of a cloud of interaction modes (photons or phonons)
around the central system. Such a cloud is already familiar from the
dressed electron picture in quantum electrodynamics.
For systems of our interest, the cloud is not present at high temperatures,
but, due to the non-vanishing coupling to the bath, it builds up at low $T$,
inducing non-thermodynamic physics.

Two level systems \cite{rmp,nmr}, like electrons or two-state
atoms in a field, or a two level Josephson junction, display
relaxational behavior due to coupling to a bath.
Transversal and longitudinal correlations relax
on time-scales $\T_2$ and $\T_1$, respectively, with
typically $\T_2\ll \T_1$ \cite{nmr}, since
energy transfer is not involved in the $\T_2$- process.

The Hamiltonian of the problem reads:
\BEA
\label{ham}
&&\H=\H_S+\H_B+\H_I,\quad \H_S=
\frac{\eps}{2}\,\si _z+\frac{\Delta}{2}\si_x, \\
&&\H_B=\sum _k\hbar\om _k\ha^{\dagger}_k\ha _k, \quad
\H_I=\frac{1}{2}\sum _kg_k(\ha _k^{\dagger}+\ha _k)\si _z.\nn
\EEA
This is a spin $\frac{1}{2}$ interacting with a bath of harmonic oscillators
(spin-boson model \cite{rmp,lu});
$\H_S$, $\H_B$ and $\H_I$ stand for the Hamiltonians of the spin,
the bath and their interaction, respectively.
$\si _x$, $\si _y$ and $\si _z=-i\si_x\si_y$ are Pauli matrices,
and $\ha _k^{\dagger}$ and $\ha _k$ are the creation and annihilation
operators of the bath oscillator with the index $k$, while the
 $g_k$ are the coupling constants.
For an electron in a magnetic field $B$, $\eps=
\bar g\mu_B B$ is the energy, with $\bar g$ the gyro-magnetic factor
and $\mu_B$ the Bohr magneton.
We shall restrict ourselves to the model with $\Delta=0$,
which is a prototype of a variety of
physical systems \cite{rmp}, and known to be exactly solvable
\cite{rmp,lu}, since $z$-component of the spin
is conserved, and with it the spin energy.
Physically it means that we restrict ourselves to times
much less than  $\T_1$.
In NMR (ESR) physics~\cite{nmr} the model represents a
spin (electron) interacting with a bath of phonons.
In quantum optics it is suitable for describing a two-level
atom interacting with a photonic bath.

Starting from general physical arguments \cite{rmp}, one typically takes the
quasi-Ohmic spectral density of the bath
\BEA
\label{e0} J(\omega)=\sum_k \frac{g_k^2}{\hbar\omega_k}
\delta(\omega_k-\omega)=\frac{g\hbar}{\pi}e^{-\om /\Gamma},
\EEA
where $g$ is the dimensionless damping constant and the exponential cuts off
the coupling at $\omega\gg \Gamma$, the maximal frequency of the bath.
As usual, the thermodynamic limit for the bath has been taken here.

Since $\Delta=0$, one has conservation of $\si_z(t)=\si_z(0)$
(in the Heisenberg picture). The dynamics of the annihilation operator
of the $k$'th bath mode reads
\BEA
\label{gegel}\label{akt=}
\ha _k(t)=
e^{-i\om _kt}\ha _k(0)+\frac{g_k\si _z}{2\hbar \om _k}\,\left(
e^{-i\om _kt}-1
\right).
\EEA
This implies
\BEA
\label{kant}
\sum _kg_k[\,\ha _k^{\dagger}(t)+\ha _k(t)\,]=\e (t)-\si_z\,G(t),
\EEA
where we denoted the quantum noise operator
\BEA \label{eta=}
&&\e  (t)=
\sum_kg_k[\ha _k^{\dagger}(0)e^{i\om _kt}+\ha _k(0)e^{-i\om _kt}],
\label{bura}
\EEA
which will act as a random force on the spin, and where
\BEA
G(t)&=&
\sum _k \frac{g^2_k}{\hbar\om _k}(1-\cos \om _kt)
=g\frac{\hbar \Gamma}{\pi}\frac{\Gamma^2 t^2}{1+\Gamma^2t^2},
\label{transilvan}
\EEA
showing that $1/\Gamma$ is the relaxation time of the bath.

{\it Separated initial state}.
To describe situations, where the spin was suddenly brought into the
contact with the bath, e.g. an electron injected into semiconductor,
atom injected into a cavity, or exciton created by external radiation,
we make the assumption that initially, at
$t=0$, the spin and the bath are in a separated state,
the latter being Gibbsian at temperature $T=1/\beta$:
$\rho (0)=\rho _{S}(0)\otimes
\exp (-\beta \H_B)/Z_B$,
where $\rho _S(0)$ is the initial density matrix of the spin.
In this situation the quantum noise is stationary and Gaussian
with average zero and time-ordered autocorrelation function:
$K_{\cal T}(t-t')=\langle \T [\e  (t) \e  (t')] \rangle$,
where ${\cal T}$ stands for the time-ordering operator and the
brackets for the trace over the initial state.
For $t>0$ it holds that
\BEQ\label{KT=}
 K_{\cal T}(t)=K(t)-i\,\hbar\dot G(t)
\equiv\hbar^2[\ddot\xi(t)-i\,\ddot G_1(t) ]
\EEQ
where an explicit calculation yields
\BEA
\label{nepal}
\xi (t)&=&\frac{g}{\pi}\,
\ln \frac{\Gamma ^2\left(1+\frac{T}{\hbar\Gamma}\right)
\,\,\sqrt{1+\Gamma^2t^2}}{\Gamma\left (1+\frac{T}{\hbar\Gamma }
-i\frac{Tt}{\hbar}\right )\Gamma \left (1+\frac{T}{\hbar\Gamma }
+i\frac{Tt}{\hbar}\right )} \\
G_1(t)&=&\frac{g}{\pi}\Gamma t-\gamma(t),\qquad\gamma(t)=
\frac{g}{\pi}\arctan\Gamma t \label{G1t=}
\EEA

The spin operators
$\si_\pm =\si _x\pm i\,\si _y$ satisfy
\BEA
\label{k3}
\dot{\hat{\sigma}}_\pm = \frac{i}{\hbar}\,
\left[\,\pm \eps +\e(t)-G(t)\,\right]\,\si_\pm
\EEA
and have, with $\omega_0=\eps/\hbar$,  the solution
\BEA
\label{k4}
&&\si_\pm (t)=\exp \left (\pm\, i\omt\,\right)\,
\hat\Pi_\pm (t,0)\,\si_\pm (0)
\\
&&\label{Pit1t0=}
\hat{\Pi}_{\pm}(t_1,t_0)\equiv e^{-i\,G_1(t_1-t_0)}\,
{\cal T}\,\exp\left[\pm \frac{i}{\hbar}\int_{t_0}^{t_1}
\,\d s\,\e (s) \right],
\EEA
Depending only on $a_k(0)$ and $a_k^\dagger(0)$,
it commutes with $\si_{x,y,x}(0)$.
Thus one gets
\BEA
\label{k7}
\langle \sip (t)\rangle =\exp \left (i\omt\right)\,\left\langle
\,\piplus(t,0)
\,\right\rangle\langle\sip (0)\rangle,
\EEA
Evaluating the time ordered product term-by-term
with help of Wick's theorem and resumming, we obtain
\BEA
\left\langle
\,\piplus(t_1,t_0)
\,\right\rangle
=\exp[-\xi(t_1-t_0)],
\label{ugar}
\EEA
The result is stationary, as it should.
Substituting (\ref{ugar}) into (\ref{k7}), the
and real and imaginary parts give
\BEA
\label{joker1}
&&\langle \si _x(t)\rangle =\left[
\cos\omt\langle\si_x(0)\rangle -
\sin\omt\langle\si_y(0)\rangle
\right]\,e^{-\xi (t)}\\
&&\langle \si _y(t)\rangle =\left[
\sin\omt\langle\si_x(0)\rangle+\cos\omt\langle\si_y(0)\rangle
\right]\,e^{-\xi (t)}
\label{joker2}
\EEA
For $t\gg 1/\Gamma$ Eq. (\ref{nepal}) brings
\BEA \label{xi=}
\xi(t)
\approx \frac{t}{\T_2},\qquad \T_2=\frac{1}{g}\,\frac{\hbar}{T}
\EEA
$\T_2$ can thus be identified with the transversal decay time.
For small $g$ this is a strong enhancement of the quantumtimescale $\hbar/T$.
The most optimistic case, $\T_2\sim 100$ s
at room temperature, involves $g\sim 10^{-16}$, truly small.

The density matrix of the spin reads
\BEA
\rho _S=\frac{1}{2}\left [\,
1+\langle \si _x(t)\rangle\, \si _x+
\langle \si _y(t)\rangle \,\si _y +\langle \si _z(t)\rangle \,\si _z\,
\right ].
\EEA
Its von Neumann entropy equals
$S_{\rm vN}=-\tr \rho_S\ln\rho_S=-p_1\ln p_1-p_2\ln p_2$, where $
p_{1,2}=\half\pm\half|\langle{\vec{\sigma}}\rangle|$.
In the course of time $|\langle\vec{\sigma}(t)\rangle|$
decays to $|\langle\si_z(0)\rangle|$, which makes
the von Neumann entropy increase.
Since there is no heat flow - the energy is conserved - this is
in agreement with a formulation of the second law: the entropy
of closed system, or of an open system without energy transfer
(the spin in contact with the bath), cannot decrease.

{\it A sudden pulse}.
We mostly consider the Hamiltonian (\ref{ham}) with $\Delta=0$.
A fast rotation around the $x$-axis is described by taking $\Delta\neq 0$
during a short time $\delta_1$; this is called a fast pulse\cite{nmr}.
If $\Delta\sim 1/\taut$ is large,
the  evolution operator describing the pulse becomes
$U_1=\exp(-i\delta_1\H(\Delta)/\hbar)\approx \exp({\half i\,\theta\si_x})
+{\cal O}(\taut)$, where $\theta$$=$$-\delta_1\Delta/\hbar$ is
the rotation angle,
\BEA
\label{h12}
&&U_1^{-1}\,\si_y\,U_1=
\si _z\sin \theta +\si _y\cos \theta ,
\\ \label{h11}
&& 
U_1^{-1}\,\si_z\,U_1
=\si _z\cos \theta -\si _y\sin \theta.\EEA
When suddenly switching $\Delta$ on and off, the state of
the system does not change,
so $\rho (t+\taut )=U_1\,\rho (t)\,U^{-1}_1$.
The work done by the source is the change of the total energy
which reads, since $[U_1,\H(\Delta)]=0$,
\BEA
\label{2}
W_1(t)&=&{\rm tr}\left[\rho(t)(\H(\Delta)-\H)+\rho (t+\taut)(\H-\H(\Delta))
\right]
\nn\\&=&{\rm tr}\,\rho(t)(U^{-1}_1\H U_1-\H),
\EEA

The work done in this variation appears to be
\BEA\label{maugli}
W_1=&-&\frac{\eps}{2}\,(1-\co)\langle\si_z(0)\rangle
-\frac{\eps}{2}\so\langle\si_y(t)\rangle \\
&+&
\half (1-\co)\,G(t)-\frac{\hbar\dot\xi(t)}{2}\so\langle\si_x(t)\rangle\nn
\EEA
Our main interest is work extraction from the bath.
In order to ensure that the pulse does not change the energy of the spin,
we first consider the case $\eps =0$, where the spin has no energy.
For small $g$, $\theta=-\pi/2$ and $ t\gg 1/\Gamma$ one gets
\BEQ\label{W1==}
 W_1=\frac{g\hbar\Gamma}{2\pi}+
\frac{gT}{2}\langle\si_x(0)\rangle\,e^{-t/\T_2}\EEQ
If for a fixed $t$, temperature is neither too large nor too small,
$Te^{-t/\T_2}> \hbar\Gamma/\pi$,
work can be extracted ($W_1<0$),  provided
the spin started in a coherent state  $\langle\si_x(0)\rangle= -1$.
This possibility to {\it extract work from the bath}
disappears on  the timescale $\T_2$, because then the spin looses its
coherence, $\langle\si_{x,y}(t)\rangle\to 0$.
Without a pulse the spin energy is conserved, however.
Notice that any combination of $\pm\pi$ pulses (this is a
classical variation, since the coherence is not involved)
can extract work only from a non-thermalized bath, i.e.
for times $\sim 1/\Gamma$.

{\it Initial preparation via a rotation.}
Our approach also allows to consider a specific, well controllable
non-equilibrium initial state: a Gibbsian of the total system,
$\rho_G=\exp(-\beta\H)/Z$, in which at $t=0$ the spin is rotated
over an angle $-\half\pi$ around the $y$-axis,
$\rho(0)=U_0\,\rho_G\,U_0^{-1}$, with $ U_0=\exp(-i\pi\si_y/4)$. 
This maps $\si_x\to\si_z$, $\si_z\to-\si_x$.
Such a state models the optical excitation of the spin, as it is done
in NMR and spintronics.
Though $\rho(0)$ does not have the product form,
the problem remains exactly solvable.
Taking $\theta=-\half\pi$ one now gets
\BEA &&W_1=\frac{G(t)}{2}-\left[
\frac{\eps}{2}(\sin\omt\cos\gamma\tanh\frac{\beta\eps}{2}
+\cos\omt\sin\gamma)\right.\nn\\&&+\left.\frac{\dot\xi(t)}{2}
(\cos\omt\cos\gamma\tanh\frac{\beta\eps}{2}-\sin\omt\sin\gamma)
\right]\,e^{-\xi(t)}\nn\\
&&\approx\frac{g\hbar\Gamma}{2\pi}-
\left[\frac{\eps}{2}\sin\omega_0t+\frac{gT}{2}\cos\omega_0t\right]
\tanh\frac{\beta\eps}{2}\,\,e^{-t/\T_2}
\EEA
where $\gamma(t)$ of Eq. (\ref{G1t=}) arises from friction,
with $\gamma(\infty)=\half g$, and where $\omega_0=\eps/\hbar$.
Typically  $g$ is small, so  work is extracted ($W_1<0$)
when the sinus is positive. The work decomposes,
\BEA W_1=\Delta U-\Delta Q\EEA
into the change in spin energy due to the pulse,
\BEA \Delta U&=&\frac{\eps}{2}
[\langle\si_z(t^+)\rangle-\langle\si_z(t^-)\rangle]
=\frac{\eps}{2}\langle\si_x(t^-)\rangle\nn\\&=&
-\frac{\eps}{2}
[\sin\omt\cos\gamma\tanh\frac{\beta\eps}{2}+
\cos\omt\sin\gamma]\,e^{-\xi(t)}\nn\\&
\approx& -\frac{\eps}{2}
\sin\omt\tanh\frac{\beta\eps}{2}\,e^{-t/\T_2},
\EEA
and the heat absorbed from the bath
\BEA \Delta Q\approx\frac{g}{2}
\left[-\frac{\hbar\Gamma}{\pi}+T\cos\omt\,\tanh\frac{\beta\eps}{2}
\,e^{-t/\T_2}\right]
\EEA
 Notice its similarity with $-W_1$ of Eq. (\ref{W1==}).
An interesting case is where work is performed by the total
system ($W_1<0$) solely due to  heat  taken from the bath ($\Delta Q>0$,
$\Delta U=0$). This process, possible by choosing $t\approx 2\pi n/\omega_0$
with integer $n$, can be considered as a cycle of a perpetuum mobile,
forbidden by folklore minded formulations of the second law.
Indeed, under a rotation the length $|\langle\vec{\sigma}\rangle|$,
and with it the von Neumann entropy, is left invariant,
so one has a process with
$  \Delta Q>0$ , $\Delta S_{\rm vN}=0$,
which violates the Clausius inequality
$\Delta Q\le T\Delta S_{\rm vN}$.

The work needed at time zero to rotate the spin is
\BEA W_0&=&-\tr \rho_G\frac{\eps+\eta_a}{2}(\si_x+\si_z)=
\frac{\eps}{2}\tanh\frac{\beta\eps}{2}+\frac{g\hbar\Gamma}{2\pi}
\EEA
representing the work done on the spin and on the bath, respectively.
It can be verified that the total work $W_0+W_1$ is always positive, so
Thomson's formulation for a cyclic change~\cite{ANthomson}
(here: the combination of the pulses at time $t=0$ and $t$)
starting from equilibrium is obeyed.

{\it Two pulses in a rotated initial Gibbsian state.}
If there are many spins, each in a slightly different external field,
there appears an inhomogeneous broadening of the $\omega_0=\eps/\hbar$
line, for which we  assume the distribution
\BEQ p(\omega_0)=
\frac{2}{\pi} \frac{[\T_2^\ast]^{-1}}{(\omega_0-\bar\omega_0)^2
+[\T_2^\ast]^{-2}}
\EEQ
having average $\bar\omega_0$ and inverse width $\T_2^\ast$,
typically much smaller than $ \T_2$.
In this case the gain for a single pulse is washed out,
leaving only the loss $\Delta Q=-g\hbar\Gamma/2\pi$,
so two pulses are needed.
We consider again the rotated initial Gibbsian state,
and perform a first $-\half\pi$ pulse around the $x$-axis
at time $t_1$ and a second $\half\pi$ pulse
at time $t_2=t_1+\tau$.
In the regime of small $g$ and large $t_1\sim \T_2$
the work in the second pulse is
\BEA
&&W_2=\frac{g\hbar\Gamma}{2\pi}-\half e^{-t_1/\T_2}
\eps\sin\omtau\,\tanh\frac{\beta\eps}{2} \\&&
-\half e^{-t_2/\T_2}\tanh\frac{\beta\eps}{2}\,\cos\omega_0t_1
(\eps\sin\omega_0\tau+gT\cos\omega_0\tau)\nn
\EEA
At moderate times only slowly oscillating terms survive. They are the ones
that involve $\Delta t=t_2-2t_1$.
For the total work $W_1+W_2$ this brings
\BEA W&=&\frac{g\hbar\Gamma}{\pi}
-\frac{\hbar}{4} e^{-t_2/\T_2}e^{-|\Delta t|/\T_2^\ast}
\tanh\frac{\beta\hbar\bar\omega_0}{2}\left
\{\bar\omega_0\sin\bar\omega_0\Delta t\right.
\nn \\&+&
[\frac{1}{\T_2}-\frac{{\rm sg}(\Delta t)}{\T_2^\ast}
(1+\frac{\beta\hbar\bar\omega_0}{\sinh\beta\hbar\bar\omega_0})]
\cos\bar\omega_0\Delta t\,\}  \EEA
For  $\Delta t$ near $2\pi n/\bar\omega_0$ such that the odd terms
cancel, this again exhibits work extracted solely from the bath.

{\it Bath-induced gain without inversion}. It is common knowledge
that a two-level system with population inversion, i.e. with a negative
temperature, is capable to amplify light,  and it represents the basic
working mechanism of lasers and masers (see \cite{opt} for more
non-standard mechanisms of lasing).
In this context a bath is typically considered to be a drawback
for the amplification,  as being a source of undesirable noises and
relaxation towards equilibrium \cite{opt}.
Our present aim is to show that the bath can nevertheless play a totally
different role, namely in {\it assisting} work extraction (gain) by means
of an {\it positive} temperature state in the two-level system.
This effect is strictly
prohibited by the second law applied to a  positive temperature spin state
if there were no coupling to the bath ~\cite{ANthomson}.
We consider separated initial conditions
with $\langle \si_x(t)\rangle=\langle \si_y(t)\rangle=0$,
and apply a $-\half\pi$ pulse at time $t_1=0^+$, and a
$\half\pi$ pulse at $t_2$.
For $t_2\gg 1/\Gamma$ the work is:
\BEA
&&W=\Delta U-\Delta Q,\\
&&\Delta U=-\frac{\eps}{2}\,[1-\,e^{-\xi(t_2)}\cos\omega_0t_2]
\langle\si_z\rangle
+\frac{g\eps}{4}\,e^{-\xi(t_2)} \sin\omega_0t_2\nn\\
&&\Delta Q=-\frac{g\hbar\Gamma}{\pi}+
\half gT\,e^{-\xi(t_2)}\sin\omega_0 t_2\langle\si_z\rangle
\EEA
where $\xi$ was defined in (\ref{xi=}).
In the inversion-free case, the initial state of the spin is
a Gibbsian connected to a positive temperature $T_0=1/\beta_0$, for which
$\langle\si_z\rangle=-\tanh\half\beta_0\eps\le 0$.
Let us first investigate the case $T_0=\infty$
(completely random state, $\langle \si_{x,y,z}\rangle=0$).
The work $W$ can be negative (gain) provided
$\eps> 4\hbar\Gamma/\pi$. This situation can be met in quantum optical
two-level systems \cite{opt,atoms} and in NMR \cite{nmr1}.
This mechanism concerns work extraction {\it with help of the bath}
(it disappears for $g\to 0$), but
{\it not from the bath}, since now $\Delta Q<0$.
The origin of the effect  is
that although the state of the spin was completely
disordered initially, the first pulse does generate some coherence.
Due to back-reaction of the bath one has  after the pulses
$\langle\si_y(t_2)\rangle=
\sin\gamma(t_2)\exp(-\xi(t_2))\sin\omega_0 t_2$,
where  $\gamma(t_2)$ of Eq. (\ref{G1t=})
goes from $0$ to $\half g$ on the timescale $1/\Gamma$,
the reaction time of the bath.

At finite $T_0$ the term $\Delta U$
can still be negative when $T_0\gtrsim \eps/g$, which can be met for
not-too-small $g$, a condition anyhow needed for having a sizeable effect.
>From a thermodynamic point of view the gain can be seen as a
flow of energy from a high temperature (of the spin) to a lower one
(of the bath), and the outside world (gain).

{\it Feasibility.} Let us present several reasons favoring the feasibility of
the proposed setups: 1) Two-level systems are widespread, not the least because
many quantum system act as two-level under proper conditions; 2) Detection
in these systems is relatively easy, since already one-time quantities
$\langle\vec{\sigma}(t)\rangle$ completely determine the state;
3) The harmonic oscillator bath is universal \cite{ms};
4) Work and heat were measured in NMR experiments more than 35 years ago
\cite{Schmidt}; 5) Our main effects do survive the averaging over disordered
ensembles of spins, thus allowing many-spin measurements.
6) The ongoing activity for implementation of quantum computers
provides experimentally realized examples of two-level systems, which
have sufficiently long ${\cal T}_2$ times,
and admit external variations on times smaller than ${\cal T}_2$:
({\it i}) for atoms in optical traps ${\cal T}_2\sim 1$s,
$1/\Gamma\sim 10^{-8}$s, and there are efficient methods for
creating non-equilibrium initial states and manipulating atoms by
external laser pulses \cite{atoms}; ({\it ii}) for an
electronic spin injected or optically excited in a semiconductor
${\cal T}_2\sim 1\,\mu$s \cite{spintronics};
({\it iii}) for an exciton created in a quantum dot
${\cal T}_2\sim 10^{-9}$s \cite{exciton} (in cases ({\it ii}) and ({\it iii})
$1/\Gamma\sim 10^{-13}$s and femtosecond ($10^{-15}$s) laser pulses are
available); ({\it iv}) in NMR physics ${\cal T}_2\sim 1-10$s
and the duration of pulses can be comparable with
$1/\Gamma\sim 1 \,\mu$s.

{\it Conclusion.} Like in the oscillator model \cite{ANprl},
the Clausius inequality can be violated in the spin-boson model,
namely by a single pulse. For many spins this typically does
not happen in the first pulse; it may occur in the second
pulse, while it is absent for the two pulses taken together,
because of intermediate entropy increase.
Work can be extracted from the equilibrated bath,
 for the single spin case by one pulse,
and for the many spin case by two pulses.
This contradicts Thomson's
formulation of the second law: no gain from cycles.
Gain is also possible from a positive temperature
(inversion-free) initial state, which may serve as a principle for
bath generated lasing and masing.
These effects are not in a conflict with the {\it equilibrium}
Thomson formulation of the second law, and are fully quantum:
they disappear at high temperatures
and they do not occur for classical pulses.

\end{multicols}
\end{document}